\documentclass[%
 reprint,
nofootinbib,
 amsmath,amssymb,
  aps,
floatfix,
]{revtex4-2}
\usepackage{cleveref}
\usepackage{graphicx}
\usepackage{dcolumn}
\usepackage{bm}
\usepackage{lipsum}

\begin{document}

\preprint{APS/123-QED}

\title{MK single pion production model}

\author{M. Kabirnezhad}
\affiliation{%
 University of Oxford
}%
 \email{minoo.kabirnezhad@physics.ox.ac.uk}

\date{\today}

\begin{abstract}
This paper presents an update to the MK model \cite{MK}, which was developed to describe single pion production in neutrino-nucleon interactions. Originally the MK model used the helicity amplitudes and the hadronic current form-factors of the Rein and Sehgal model \cite{RS}. The update includes a new definition for the helicity amplitudes in the first and second resonance regions, and new vector-current form factors. Fits to electron-proton scattering data were used to determine these vector-current form factors, and to assign errors to the constrained free parameters of the model. 
\end{abstract}

\maketitle


\section{Introduction}
Neutrino interactions that produce a single pion in the final state are of critical importance to accelerator-based neutrino experiments. Single pion production (SPP) channels make up the largest fraction of the inclusive neutrino-nucleus cross section in the $1-3$ GeV neutrino energy region covered by most accelerator-based neutrino beams. \\

Models of the SPP cross section processes are required to accurately predict the number and topology of observed final state particles in neutrino interactions, and to help establish the relationship between neutrino energy and energy deposition in a neutrino detector.
This includes the model of the neutrino-nucleon interaction \cite{MK,RS,Rein,Adler,HNV,sato,Luis}, and the model of the nuclear system \cite{RMF,Raul} in which the nucleon resides and which must be traversed by the interaction final-state particles.\\

The first generation of neutrino SPP models \cite{RS,Rein,Adler} were developed with the statistical uncertainties of contemporary data sets in minds and thus aimed for precision at the $10\%$ level. These models neglected contributions to the cross section or simplified the processes that were thought to contribute at the few percent level. These contributions include lepton mass, subdominant diagrams for nonresonant interactions, and resonance/nonresonant interference terms as well as the use of simple form-factors in hadronic current. The Rein-Sehgal (RS) SPP model~\cite{RS} in the NEUT \cite{NEUT} and GENIE \cite{GENIE} generators is a well known and often used as an example of these models. As neutrino physics enters the precision era these models must be updated and improved. An update for the MK model is introduced in Section \ref{update}.\\

The most onerous challenge in developing a high precision SPP neutrino interaction model is correctly including the effects of the overlapping structures of the many resonances that contribute to the hadronic tensor. Any model must contain free parameters associated with each resonance, which can be constrained by data. The free parameters of the vector current can be fitted to existing electron scattering data as was performed by Lalakulich and Paschos (LP) \cite{Olga_fit} and for the DCC model \cite{DCC}. The treatment described in this paper also includes estimates of parameter uncertainties and their propagation to the uncertainty on the full MK model, not included in \cite{Olga_fit}\cite{DCC}. \\

In this work, three distinct features are extracted from the data via fits: 1) the vector form factor of the resonances used in the resonant interaction model, 2) the nucleon form factors for the nonresonant interactions, and 3) the interference phases between resonant and nonresonant helicity amplitudes. 
\section{The update of MK model \label{update}}
Weak interaction nucleon vertices produce a single pion through two distinct channels in the invariant mass ($W< 2$ GeV) region. \emph{Resonance production} is where the exchange boson has the requisite four-momentum to excite the target nucleon to a resonance state, which then promptly decays to produce a final-state meson. In \emph{nonresonant production} the pion is created at the interaction vertex \cite{HNV}, which produces final states identical to resonance production resulting in non-negligible interference terms. \\

The MK model provides a full kinematic description of SPP in neutrino-nucleon interactions, including the resonant and the nonresonant interactions in the helicity basis. This allows for calculations of the interference terms. Resonance production in the MK model follows the formalisms of the RS model however the original dipole form-factors are replaced by Graczyk-Sobczyk form-factors \cite{GS}. The nonresonant interactions follow the HNV model \cite{HNV} which is based on chiral symmetry and it is not reliable at high hadron invariant mass (W). Therefore, in the original MK model a virtual form-factor ($F_{vir}$) was introduced to eliminate the nonresonant contributions smoothly across the $W\in[1.4-1.6] $ GeV region.\\

In the updated approach the vector helicity amplitudes from the RS model are changed to the helicity amplitudes of the Rarita-Schwinger formalism\footnote{A similar approach is used in \cite{GS}}\cite{Olga_fit}. However, the vector form-factors are defined differently to improve agreement with exclusive electron scattering data. \\

The vector helicity amplitudes in the MK model for resonances are related to $f^V_{-1}, f^V_{-3}$ and $f^V_{0+}$ as presented in the \Cref{res_HV}. The new vector helicity amplitudes for the $P_{33}(1232)$, $P_{11}(1440)$, $D_{13}(1520)$, and $S_{11}(1535)$ resonances are the following:
\begin{itemize}
\item{Resonance $P_{33}(1232)$}
\begin{eqnarray} \label{p33}
  f^{V}_{-3} &= &-\frac{|{\bf{k}}|}{\sqrt{2W(E_k +M)} }\Big[\frac{C_3 W_{+}}{M} \nonumber\\
  &+& \frac{C_4}{2M^2} (W_{+}W_{-}+k^2 ) +  \frac{C_5}{2M^2} (W_{+}W_{-} -k^2 )\Big],\nonumber\\
 f^{V}_{-1} &= &-\frac{|{\bf{k}}|}{\sqrt{6W(E_k +M)} }\Big[\frac{C_3}{MW}\left(k^2 - MW_{+}\right) \nonumber\\
  &+& \frac{C_4}{2M^2} (W_{+}W_{-}+k^2 ) +  \frac{C_5}{2M^2} (W_{+}W_{-} -k^2)\Big],\nonumber\\
f^{V}_{0+} &= &\frac{\sqrt{-k^2}}{M}\frac{|{\bf{k}}|}{\sqrt{3W(E_k +M)} }\Big[\frac{C_3}{M}W  + \frac{C_4}{M^2}W^2\nonumber\\
  &+&   \frac{C_5}{2M^2} (W^2 + M^2 - k^2)\Big]
 \end{eqnarray} 
 \item{Resonance $D_{13}(1520)$}
 \begin{eqnarray} 
  f^{V}_{-3} &= &\sqrt{\frac{E_k + M}{2W}}\Bigg[\frac{C_3}{M}W_{-} + \frac{C_4}{2M^2} (W_{+}W_{-} +k^2) \nonumber\\
  &+&  \frac{C_5}{2M^2} \left(W_{+}W_{-} - k^2\right)\Bigg],\nonumber\\
   f^{V}_{-1} &= &\sqrt{\frac{E_k + M}{6W}}\Bigg[\frac{C_3}{M}\left(W _{-}- 2\frac{{\bf{k}}^2}{E_k + M} \right) \nonumber\\
 &+& \frac{C_4}{2M^2}\left(W_{+}W_{-} + k^2\right)  +  \frac{C_5}{2M^2}\left(W_{+}W_{-} - k^2\right)  \Bigg],\nonumber\\
f^{V}_{0+} &= &\frac{\sqrt{-k^2}}{M}\sqrt{\frac{E_k + M}{3W}} \Bigg[-\frac{C_3}{M}W - \frac{C_4}{M^2}W^2\nonumber\\
  &+&   \frac{C_5}{2M} \left(W^2 +M^2 -k^2\right)\Bigg]
 \end{eqnarray} 
\item{Resonance $P_{11}(1440)$ }
  \begin{eqnarray}
  f^{V}_{-1} &= &\frac{|{\bf{k}}|}{\sqrt{W(E_k +M)}} \left[g_1 - \frac{g_2}{W_{+}^2} k^2\right] \nonumber\\
f^{V}_{0+} &= &-\frac{W}{M}\frac{|{\bf{k}|}\sqrt{-k^2}}{\sqrt{2W(E_k +M)} }\frac{1}{W_{+}}\left[ g_1 - g_2\right]
 \end{eqnarray} 
 \item{Resonance $S_{11}(1535)$}
\begin{eqnarray}\label{s11}
  f^{V}_{-1} &= &\sqrt{\frac{(E_k + M)}{W}}\left[\frac{g_1}{W_{+}^2} k^2 - \frac{g_2}{W_{+}}W_{-} \right] \nonumber\\
f^{V}_{0+} &= &\frac{\sqrt{-k^2}}{|{\bf{k}}|}\frac{W}{M}\sqrt{\frac{(E_k + M)}{2W}}\left[-\frac{g_1W_{-}}{W_{+}^2} + \frac{g_2}{W_{+}}\right] \end{eqnarray} 
 \end{itemize}
   where $E_k=\sqrt{M^2 + {\bf{k}} ^2}$ and $W_{\pm} = W\pm M$\footnote{Notations mirrors that of the original MK model paper \cite{MK}.}.
 $C_3$, $C_4$ and $C_5$ ($g_1$ and $g_2$) are form-factors for resonances with spin 3/2 (1/2), which are extracted from electron scattering fit.\\
 
For the higher mass resonances the RS helicity amplitudes with dipole form factors are used. The nonresonant interaction model is unchanged from the original MK model, but the proton form-factors are defined differently as a result of the fit as described below in the Sec. \ref{analys}. An adjustable phase between resonances and nonresonant helicity amplitudes is also included in the fit along with the parameters that govern the form factors. The results of the fit to SPP electron-proton scattering data are shown in the Sec. \ref{analys}.
\section{Analysis of electron-induced exclusive data\label{analys}}
Exclusive charged-current SPP electron-proton scattering data are used to fit the relevant free parameters of the model. The standard cross-section formula for the single pion electro-production is the following: 
\hspace*{-0.5cm}\vbox{\begin{align}
 &\frac{d^5 \sigma_{ep \rightarrow e' \pi N}}{dE_{e'} d\Omega_{e'} d\Omega^{\ast} _{\pi}}\nonumber\\
& =\Gamma_{em} \Big[ \frac{d\sigma_T}{d\Omega^{\ast} _{\pi} }   + \epsilon \frac{d\sigma_L}{d\Omega^{\ast} _{\pi} }  + \sqrt{2\epsilon(1+ \epsilon)} \frac{d\sigma_{LT}}{d\Omega^{\ast} _{\pi} } \cos \phi^{\ast}_{\pi}\nonumber\\
  & + \epsilon \frac{d\sigma_{TT}}{d\Omega^{\ast} _{\pi} }\cos 2\phi^{\ast}_{\pi}  + h_e \sqrt{2\epsilon(1+ \epsilon)} \frac{d\sigma_{LT'}}{d\Omega^{\ast} _{\pi} } \sin \phi^{\ast}_{\pi}  \Big].
 \label{em_Xsec}
  \end{align}}  
Pion angles, $\Omega^{\ast} _{\pi}$, are defined in the resonance rest frame, where
  \begin{eqnarray}
  \Gamma_{em} = \frac{\alpha}{2\pi^2 Q^2} \frac{E_{e}}{E_{e'}} \frac{q_{\gamma}}{1-\epsilon} 
  \end{eqnarray}  
is the virtual photon flux factor with $q_{\gamma}=(W^2 - M^2)/2M$, and $\epsilon = [1 + 2(q_{\gamma}^2 /Q^2) \tan^2 (\delta/2)]^{-1}$ where $\delta$ is the scattering angle of electron. \\

Setting $\Gamma_{em}$ to unity in Eq. (\ref{em_Xsec}) recovers the neutrino SPP differential cross-section formula. The SPP differential cross-section in the MK model is given in terms of the Lorentz invariants $W$ and $Q$ such that it is expressed by the outgoing lepton kinematics via:
 \begin{eqnarray}
 \frac{d \sigma_{(lp \rightarrow l' \pi N)}}{dE_{e'} d\Omega_{e'} } = \frac{ME_{e}E_{e'}}{\pi W}\frac{d \sigma_{(lp \rightarrow l' \pi N)}}{dW dQ^2 },
 \label{xsec_kin}
 \end{eqnarray}  
where $E_{e}$ ($E_{e'}$) are the incoming (outgoing) lepton energy in the lab frame.\\

  To predict the electron-nucleon interaction in the MK model, Eqs. (\ref{em_Xsec}) - (\ref{xsec_kin}) are used. Only the helicity amplitudes given in \Cref{res_HV} are used to construct the hadronic current. The vector helicity amplitudes are related to vector form factor via $f^V_{-1}$, $f^V_{-3}$ and $f^V_{0+}$ of Eq. (\ref{p33}) - (\ref{s11}).\\

Fits were used to determine the $Q^2$ dependence of the transition form-factors for resonance production and nonresonant SPP. Measurements of the single pion differential cross sections of electron scattering off a Hydrogen target collected and analysed by the CLAS Collaboration analysis is used in the fits \cite{pi0,pi+}. Data was limited to the kinematic region most important for accelerator-based neutrino experiments. The list of resonances with vector currents used in the MK model are shown in \Cref{res_list} over the kinematic ranges summarised in \Cref{clas_data}. With the limited data available ($W<1.68$ GeV) only the first seven resonances are included in the fit.\\

\begin{table}
\centering
\caption{Nucleon-resonances   }
\label{res_list}
\renewcommand{\arraystretch}{1.3}
\begin{ruledtabular}
 \begin{tabular}{lccccl}
 Resonance & $M_R[\text{MeV}]$ &$\Gamma_0[\text{MeV}]$& $\chi_E$&$ \sigma^D$ &phase  
 \\ [0.1ex]
 \hline
 $P_{33}(1232)$ & 1232 & 117& 0.994   & + & 3.52\\
$ P_{11}(1440)$ & 1440 & 350& 0.65& -  &2.73\\
 $D_{13}(1520)$ &1515  & 115& 0.60& + &2.99\\
 $S_{11}(1535)$ & 1530&150 &  0.45& -  &2.60\\
 $P_{33}(1600)$ & 1570 & 250& 0.16   & +&---\\
 $S_{31}(1620)$ & 1610 &130& 0.3 & - &--- \\
  $F_{15}(1680)$&1685&120 & 0.65  & - &--- \\
  $D_{33}(1700)$&1710&300& 0.15   & + &--- \\
$  P_{11}(1710)$&1710&140& 0.11   & -  &---\\
$  P_{13}(1720)$ & 1720&250& 0.11 & + &--- \\
$ F_{35} (1905)$ & 1880&330& 0.12 & -  &---\\
$ P_{31} (1910)$ & 1900&300 &0.22 & - &---\\
$ P_{33}(1920)$ &  1920&300 & 0.12& + &---\\
$ F_{37}(1950)$ & 1930&285 &0.40  & + &---\\
\end{tabular}
\end{ruledtabular}
\end{table}
\begin{table}
\centering
\caption{Data Kinematic Range}
\label{clas_data}
\renewcommand{\arraystretch}{1.3}
\begin{ruledtabular}
 \begin{tabular}{lccl}
 Channel & $E_e$ (GeV)&$Q^2$ Range $(\text{GeV}/c)^2$& $W$ Range (GeV) 
 \\ [0.1ex]
 \hline
$e p \rightarrow e p \pi^{0}$ & 1.645, 2.445  & 0.4 - 0.9  & 1.1 - 1.68\\
$e p \rightarrow e n \pi^{+}$ & 1.515  & 0.3 - 0.6  & 1.11 - 1.57\\
\end{tabular}
\end{ruledtabular}
\end{table}
 The fits of the various resonance are done using a bootstrapping approach. The $Q^2$ dependence of the resonance $P_{33}(1232)$  (along with the proton for the nonresonant interactions) is fit to data with $W<1.28$ GeV. The results of this first fit are then included with data of $W<1.44$ GeV to extract the $P_{11}(1440)$ resonance $Q^2$ dependence and the corresponding virtual form-factor. This process is repeated twice more with data up to $W<1.54$ GeV used to extract $D_{13}(1520)$ and $S_{11}(1535)$, and the data up to $W<1.68$ GeV to extract $P_{33}(1600)$, $S_{31}(1620)$ and $F_{15}(1680)$. \\
 
 Fits were performed for a variety of possible form-factors as defined in \cite{Olga_fit}, with additional tweaks to the exact functional form and to the number of free parameters. The form factors that resulted in fits with the lowest $\chi^2$ and stable minimum were selected, and can be found in Eqs. (\ref{C_FF}) - (\ref{nonres}). 
 In the final step all the parameters in the form-factors and the phases between these resonances and the nonresonant helicity amplitudes were fit. The result can be found in Tables \ref{res_list} and \ref{res_fit}.\\
 
\begin{figure*}
\centering
 \includegraphics[width=\textwidth]{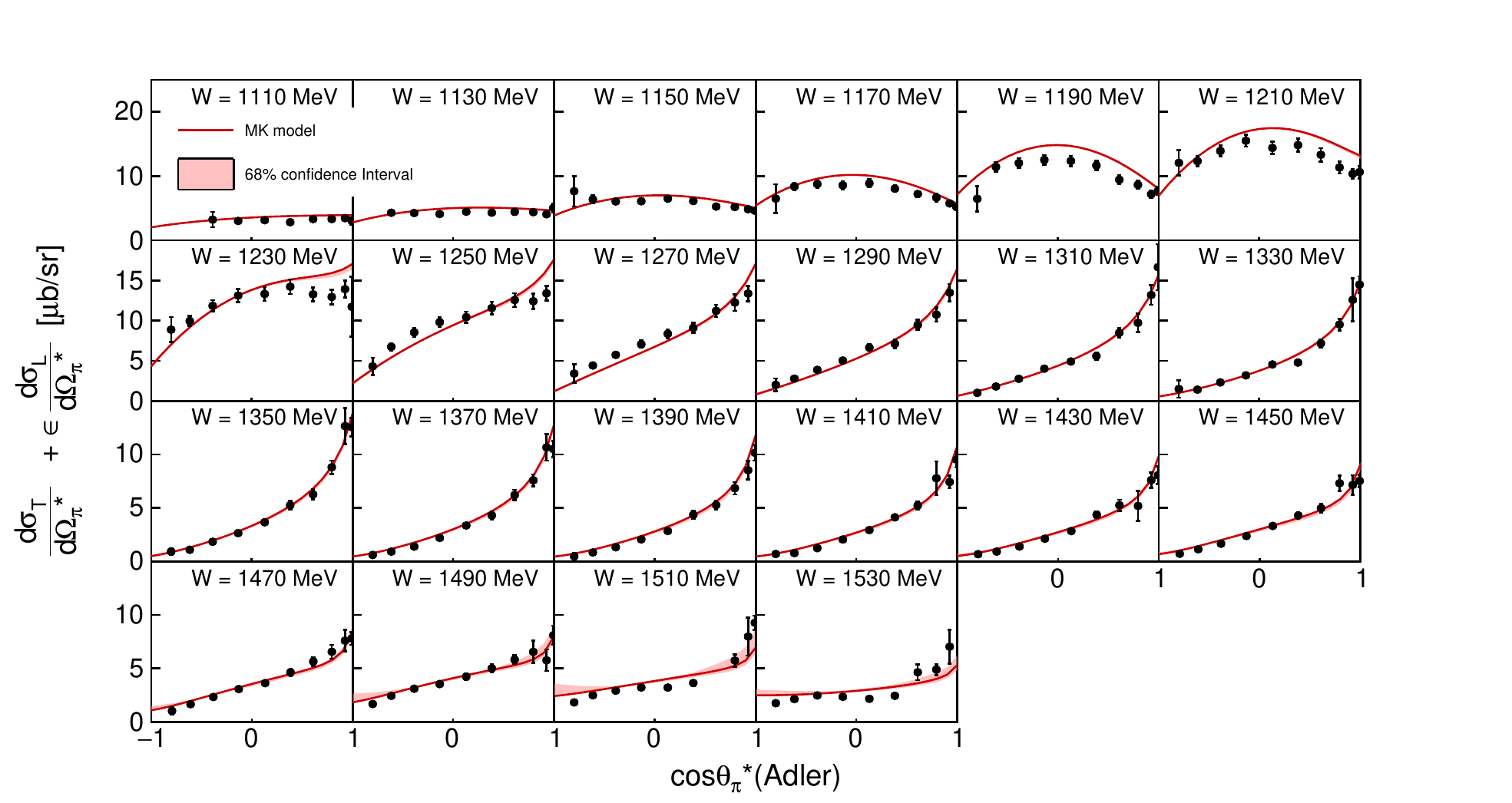}
  \caption{Fit results for $\frac{d\sigma_T}{d\Omega^{\ast} _{\pi} }   + \epsilon \frac{d\sigma_L}{d\Omega^{\ast} _{\pi} } $ at $Q^2= 0.4 ~ (\text{GeV}/c)^2$ and $E_e= 1.515$ GeV for $e p \rightarrow e n \pi^{+}$  channel from the MK model. Best values are shown as the solid red line with $68\%$ confidence interval.}
  \label{npip_0p4}
\end{figure*}

\begin{figure*}
\centering
\includegraphics[width=\textwidth]{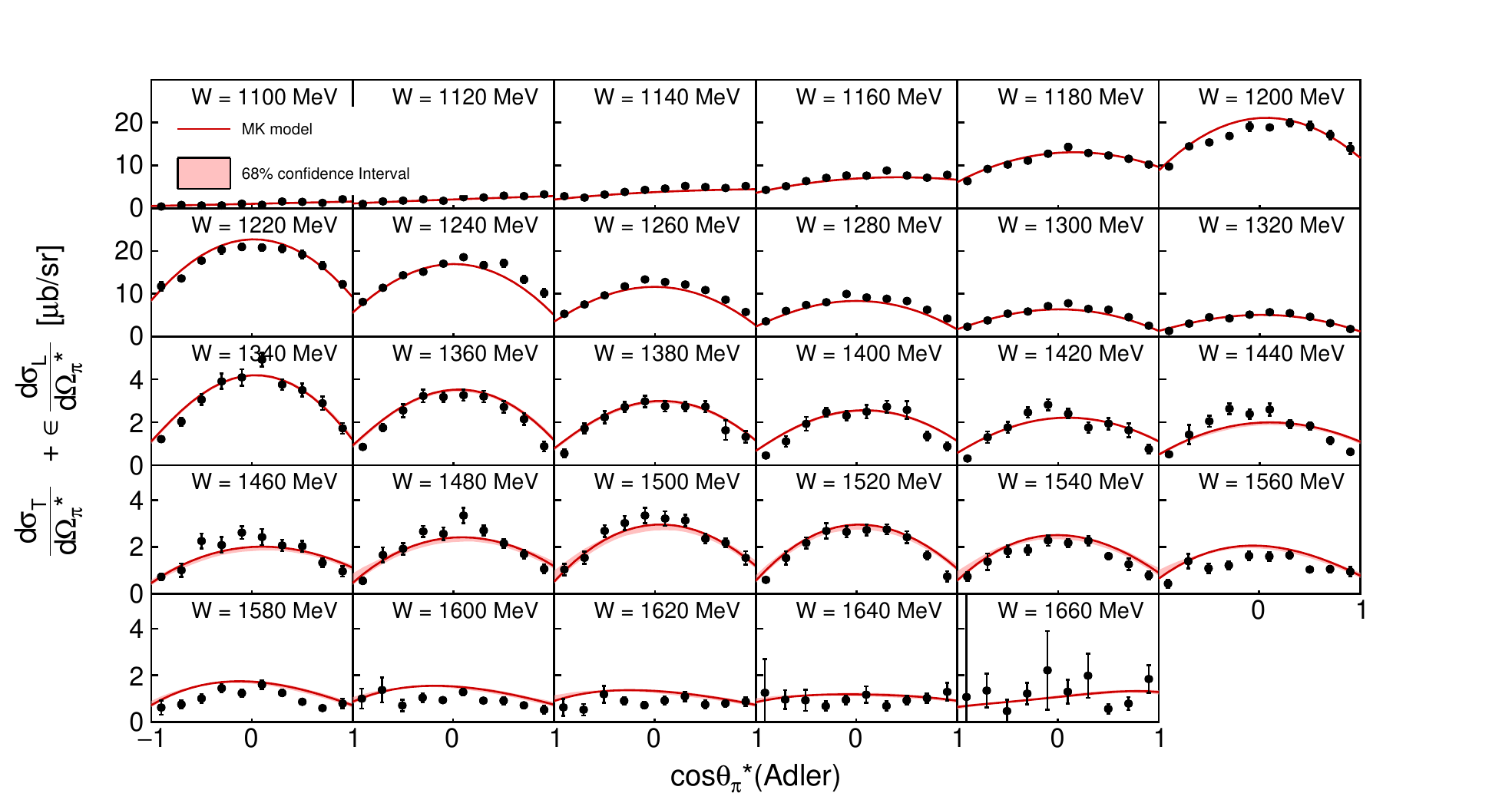}
  \caption{Fit results for $\frac{d\sigma_T}{d\Omega^{\ast} _{\pi} }   + \epsilon \frac{d\sigma_L}{d\Omega^{\ast} _{\pi} } $ at $Q^2= 0.52 ~ (\text{GeV}/c)^2$ and $E_e= 1.645$ GeV for $e p \rightarrow e p \pi^{0}$  channel from the MK model. Best values are shown as the solid red line with $68\%$ confidence interval.}
  \label{ppi0_0p52}
\end{figure*}

\begin{figure*}
\centering
\includegraphics[width=1.05\textwidth]{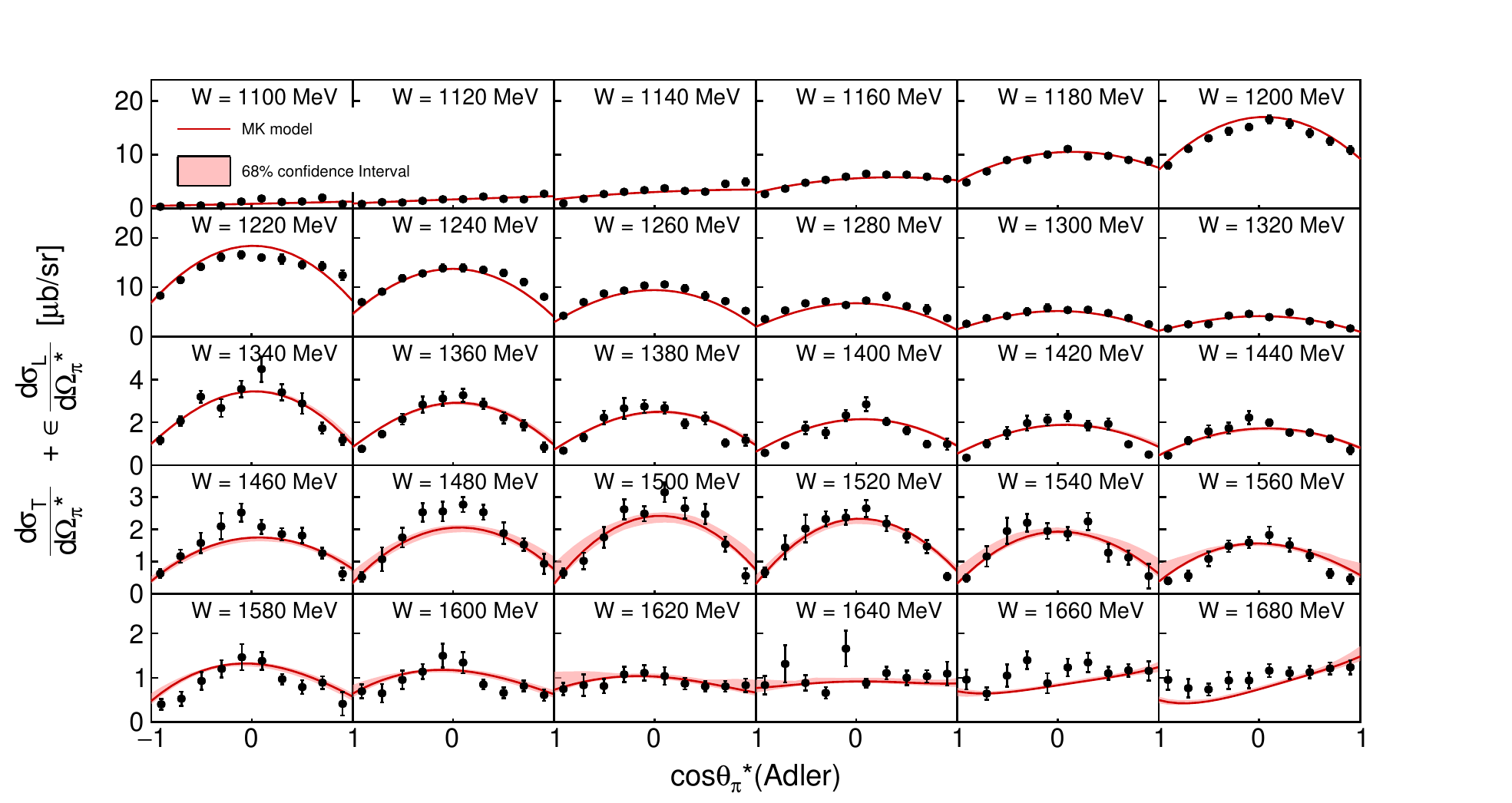}
  \caption{Fit results for $\frac{d\sigma_T}{d\Omega^{\ast} _{\pi} }   + \epsilon \frac{d\sigma_L}{d\Omega^{\ast} _{\pi} } $ at $Q^2= 0.65 ~ (\text{GeV}/c)^2$ and $E_e= 2.445$ GeV for $e p \rightarrow e p \pi^{0}$  channel from the MK model. Best values are shown as the solid red line with $68\%$ confidence interval.}
  \label{ppi0_0p65}
\end{figure*}

\begin{figure*}
\centering
\includegraphics[width=1.05\textwidth]{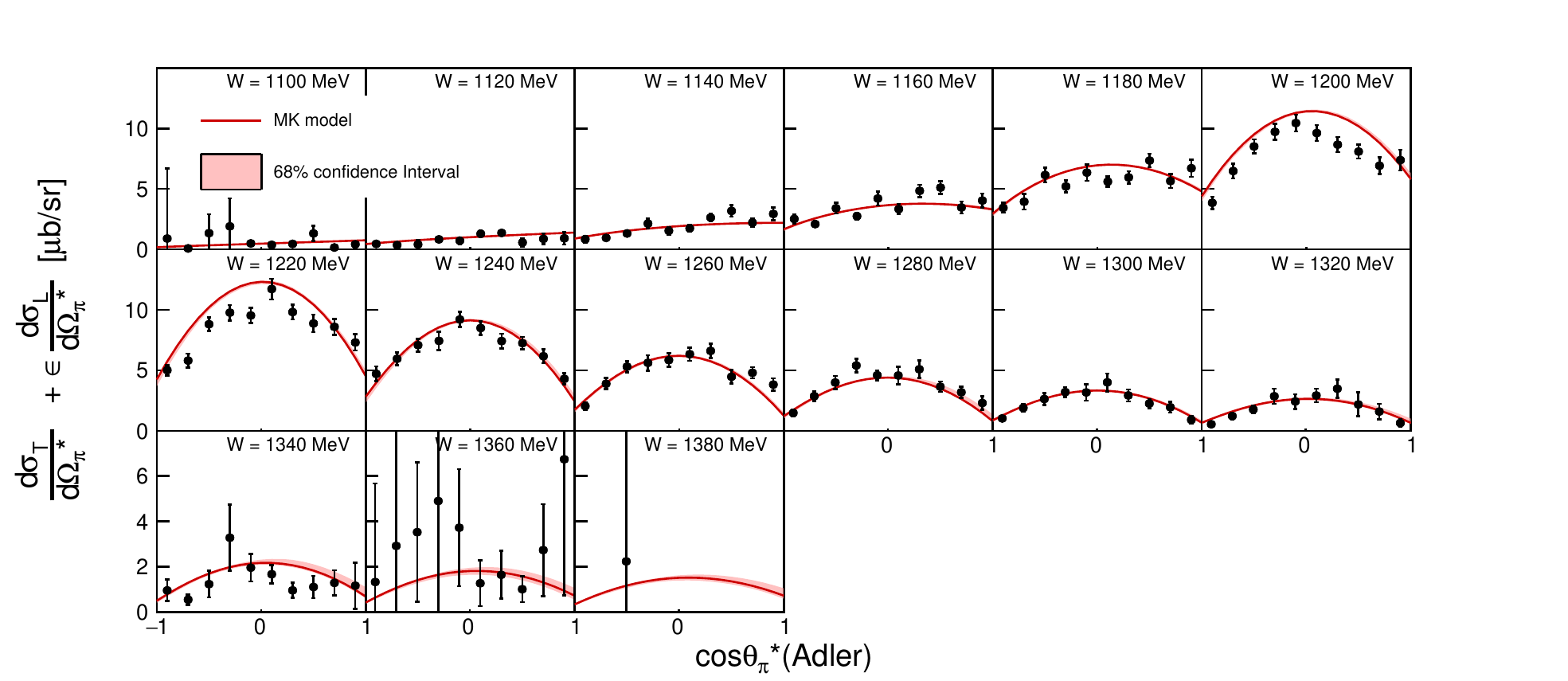}
  \caption{Fit results for $\frac{d\sigma_T}{d\Omega^{\ast} _{\pi} }   + \epsilon \frac{d\sigma_L}{d\Omega^{\ast} _{\pi} } $ at $Q^2= 0.9 ~ (\text{GeV}/c)^2$ and $E_e= 1.645$ GeV for $e p \rightarrow e p \pi^{0}$  channel from the MK model. Best values are shown as the solid red line with $68\%$ confidence interval.}
  \label{ppi0_0p9}
\end{figure*}

 \Cref{npip_0p4,ppi0_0p52,ppi0_0p65,ppi0_0p9} show some of the fit results for different $Q^2=0.4$, 0.52, 0.65, 0.9 GeV, with exclusive electron-induced reactions data within $1\sigma$ error band, where the reduced $\chi^2=2.32$. \\
     
      The fit results show that the best form factors for the resonances with spin-3/2 are:
 \begin{eqnarray} \label{C_FF}
 C_3 ^{(p)}&=& \frac{A}{(1 -k^2/M_V^2)^2}\frac{1}{1 -k^2/DM_V^2 }\nonumber\\
      C_4 ^{(p)}&= &\frac{B}{(1 -k^2/M_V^2)^2}\frac{1}{1 -k^2/DM_V^2 }\nonumber\\
      C_5^{(p)}&=& \frac{C}{(1 -k^2/M_V^2)^2} ,
 \end{eqnarray}
 while for resonances with spin-1/2 the following form factors are proposed:
  \begin{eqnarray}
g_1^{(p)}&=&\frac{A}{(1 -k^2/M_V^2)^2}\left[1 - B \ln \left(1 - \frac{k^2}{1 \text{GeV}^2}\right) \right]\nonumber\\
g_2^{(p)}&=&\frac{C}{(1 -k^2/M_V^2)^2}.
	 \end{eqnarray} 
 where $M_V= 0.84$. The best fit form-factor for resonances $P_{33}(1600)$, $S_{31}(1620)$ and $F_{15}(1680)$ are:
\begin{eqnarray}
 F_V &=& A\left( 1- \frac{k^2}{M_V^2}\right)^{-2} \left(1 - \frac{k^2}{M^2}\right)^{n/2}.
\end{eqnarray}	 
In all defined form factors, A B C and D are the free parameters that were fit to the data and their best values are listed in \Cref{res_fit}. \\

 \begin{table}
\centering
\caption{Fit result for resonance form factor}
\label{res_fit}
\renewcommand{\arraystretch}{1.3}
\begin{ruledtabular}
 \begin{tabular}{lcccl}
 Resonance & A&B&C&D
 \\ [0.1ex]
 \hline
 $P_{33}(1232)$ & -2.028 & 1.813& -0.17 & 4.73\\
$ P_{11}(1440)$ & 1.685 & 2.547& 0.494 &---\\
 $D_{13}(1520)$ &-3.689 & 3.687& -0.72&1.065\\
 $S_{11}(1535)$ &2.529&1.093 &  0.449 &---\\
 $P_{33}(1600)$ & 0.807&---&---&---\\
 $S_{31}(1620)$ & 1.510&---&---&---\\
 $F_{15}(1680)$ & 1.155&---&---&---\\
\end{tabular}
\end{ruledtabular}
\end{table}
\begin{figure*}
\centering
\includegraphics[width=1.05\textwidth]{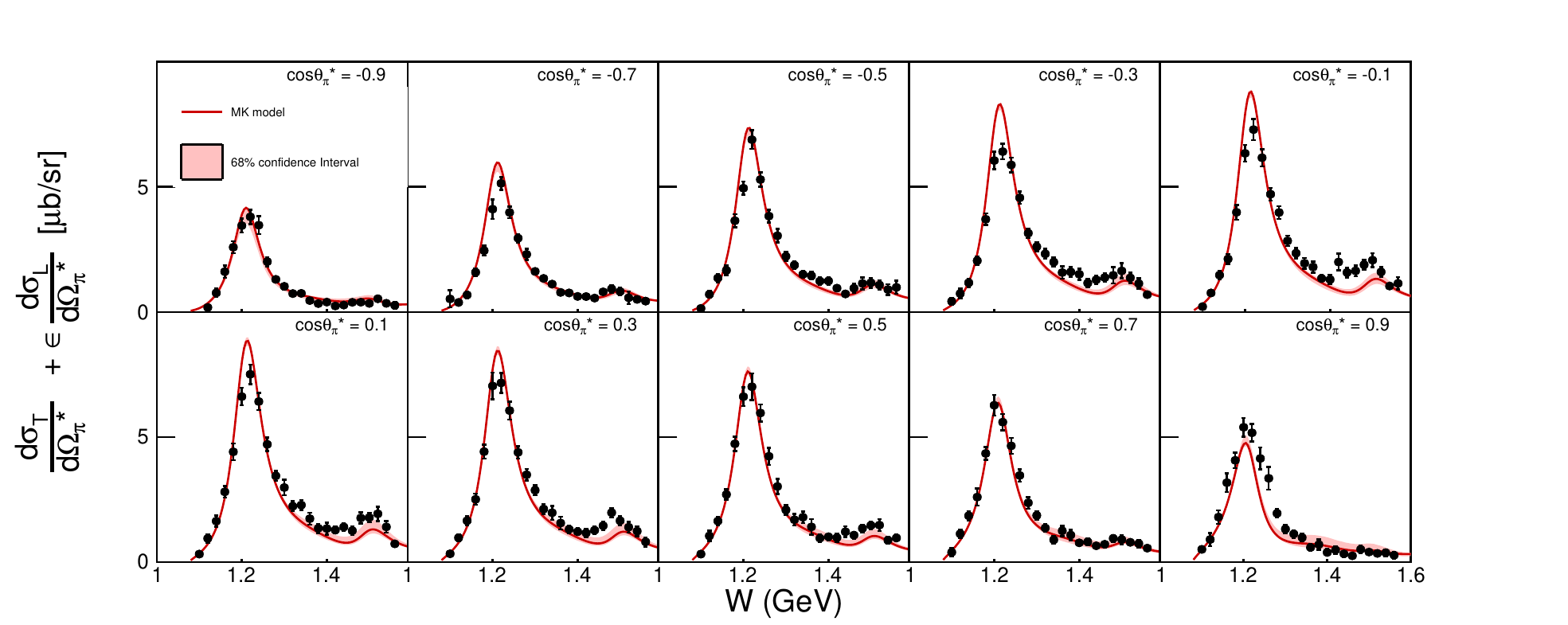}
  \caption{Comparison of the MK model with electron scattering data for different $\cos\theta^{\ast}_{\pi}$ in the  at $Q^2= 1.15 ~ (\text{GeV}/c)^2$ and $E_e= 2.445$ GeV for $e p \rightarrow e p \pi^{0}$ channel. The predictions of MK model are shown as the solid red lines with $68\%$ confidence interval and the reduced $\chi^2= 3.979$. Date is from reference \cite{pi0}.}\label{ppi0_1p15}
\end{figure*} 
The proposed form factors for the proton in  nonresonant interactions are very similar to the one in the original MK model \cite{Gal}:   
\begin{eqnarray}
F_1^{(p)} &= &\frac{1}{1+\tau}\left[ 1 + \frac{\tau}{1 + \lambda_n\tau} \left( 0.59+ \lambda_n\mu_p \tau  \right)\right] G_E\nonumber\\
\mu_p F_2^{(p)} &= &\frac{1}{1+\tau}\left[ \mu_p -1 + \frac{2.2\tau}{1 + \lambda_n\tau} \right]G_E
 \end{eqnarray} 
  with two adjustable parameters $\beta_1$ and $\beta_2$ in:
 \begin{eqnarray} \label{nonres}
  G_E&=& \left( \frac{1}{1 - k^2/ \beta_1 M_V^2} \right)^2 \nonumber\\
 \tau&=& -\beta_2 k^2/4M^2
 \end{eqnarray}
   where $\beta_1= 0.896$ and $\beta_2= 2.15$\footnote{    
     $F_{vir}(W)= 8.09W^3  -41.68W^2 + 66.34W -32.5733$ (for $1.3\text{ GeV}\leq W < 1.6\text{ GeV}$)}.
\section{Results for electron-induced data}
\Cref{ppi0_1p15} shows the MK model predictions for different pion angles and hadronic invariant mass at $Q^2= 1.15 ~ (\text{GeV}/c)^2$ where it is outside the kinematic region of the fitted data in our analysis ($Q^2< 0.9 ~ (\text{GeV}/c)^2$). The data shows the single $\pi^0$ production measurement for the different pion polar angles (in the hadronic rest frame) in terms of the invariant mass. \\

The MK model prediction can also be compared with inclusive electron scattering data. However it is important to note that the MK model can only predict a single pion in the final state while the data is the measurements of one or more pions in the final state. Therefore they are not identical, however at low ($\omega$, $W$) only a single pion can be produced in the final states, for example when $E_e<1$ GeV or when $W<1.4$ GeV. \\

\begin{figure}[h]
\centering
\includegraphics[width=0.47\textwidth]{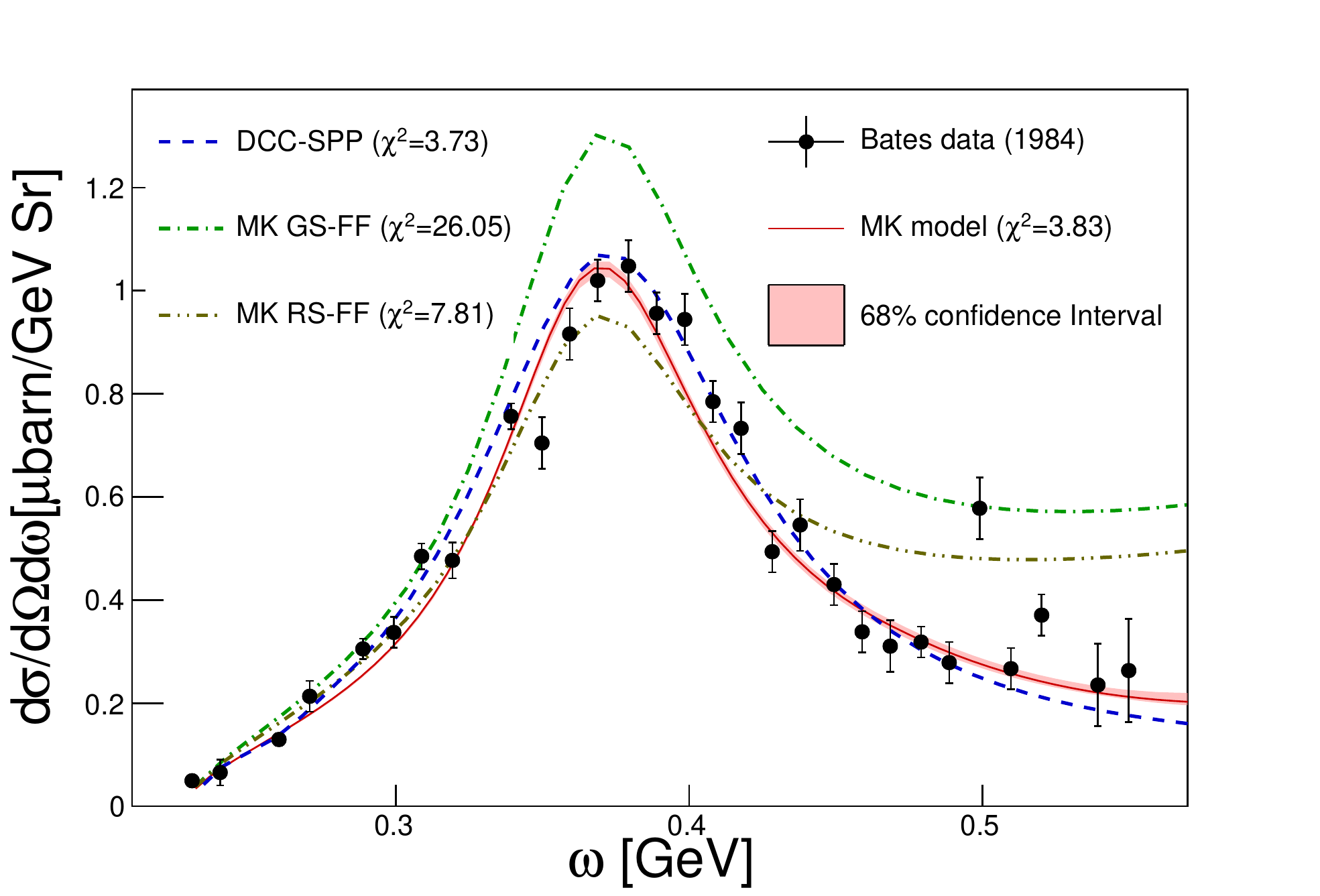}
  \caption{Inclusive differential cross section data in terms of energy transferred in the lab frame from reference \cite{bates} at $E_e= 0.73$ GeV and $\theta_e=37.1^{\circ}$ ($Q^2 \sim 0.11 (\text{GeV}/c)^2$). Updated MK model shown as the solid red lines with $68\%$confidence interval while the original MK model with GS form-factors (GS-FF) and RS form-factors (RS-FF) are shown in dashed-dotted lines. DCC model \cite{DCC} prediction for SPP is showed with dashed blue line.}
    \label{bates}
\end{figure}

\begin{figure}[h]
\centering
\includegraphics[width=0.47\textwidth]{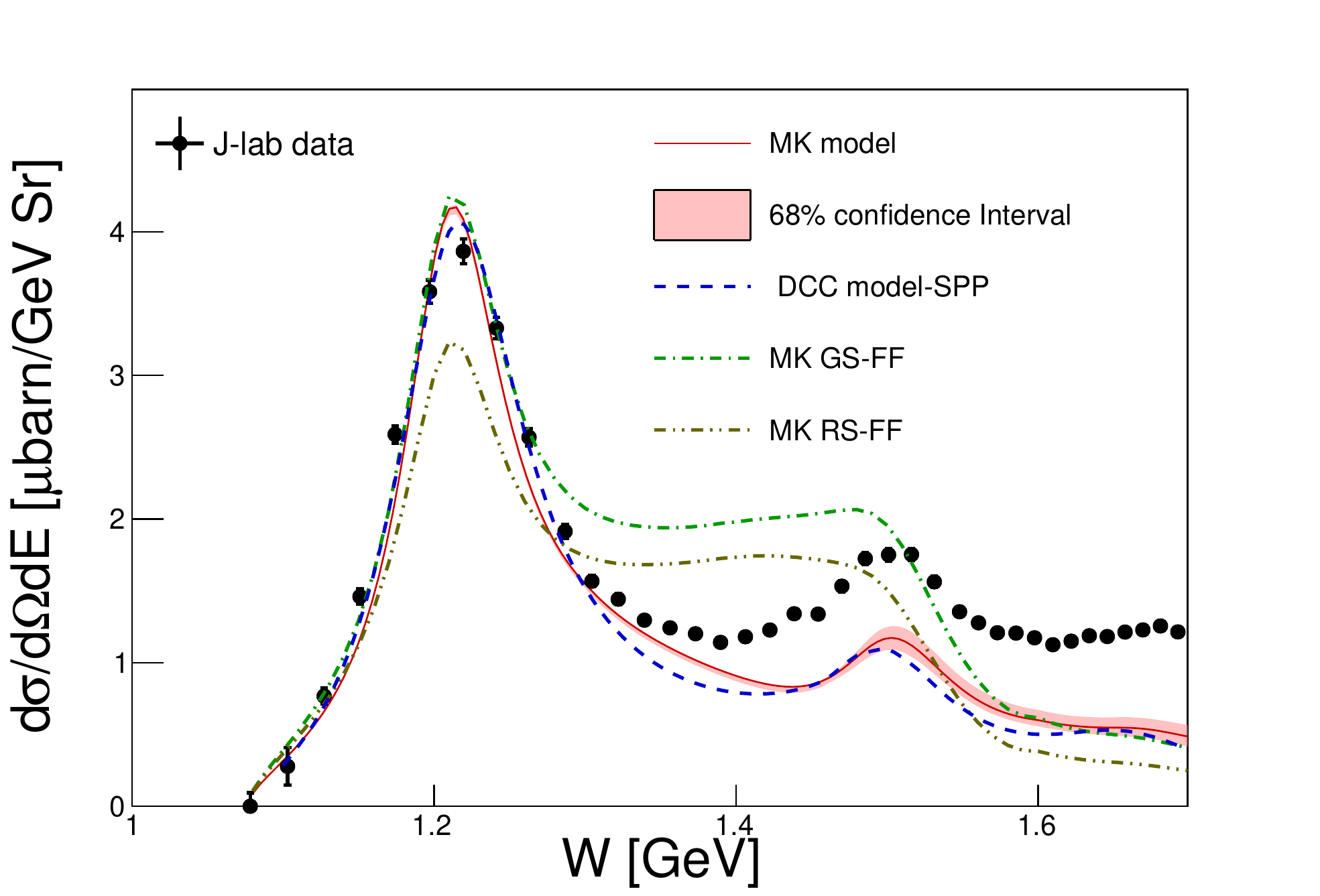}
  \caption{Inclusive differential cross section data from \cite{Jlab} at $E_e= 3.04$ GeV and $\theta_e=12^{\circ}$ ($Q^2 \sim 0.17-0.36 (\text{GeV}/c)^2$). Updated MK model shown as the red lines with $68\%$ confidence interval while the original MK model with GS form-factors (GS-FF) and RS form-factors (RS-FF) are shown in dashed-dotted lines. DCC model \cite{DCC} prediction for SPP is showed with dashed blue line.}
  \label{jlab}
\end{figure}

\Cref{bates} shows the inclusive electron-proton differential cross section in terms of energy transferred ($\omega$) at $E_e=0.73$ GeV, where exact agreement with data is expected. On the other hand, Fig. \ref{jlab} shows the same measurement at higher energy ($E_e=3.04$ GeV) where  good agreements is expected in the $\Delta$ region. To show the MK model improvements, the predictions of original MK model with the RS and the GS form-factors are displayed. The DCC model prediction for SPP are also presented. In \Cref{bates} where the data and the models are identical, the reduced $\chi^2$ are added to the legend which shows the MK and the DCC models have similar agreements with data. This also indicates that the MK model is significantly improved as $\chi^2$ is reduced from 26.05 to 3.83.
\section{conclusion}
In this work the vector current form-factors of the MK model were improved by introducing new helicity amplitudes for resonances in the first and the second resonance regions. The free parameters of the model are fit to exclusive electron scattering data to constrain the vector form factors of resonances (up to $W=1.68$ GeV) and the proton in the nonresonant interaction and evaluated the total uncertainty on cross-sections prediction. \\

 The goodness-of-fit of the model to the data shows that the pre-fit model covers the data. Further validation of the constrained model against inclusive electron scattering data sets demonstrates the robustness of the constrained model and its predictive power.
\begin{acknowledgments}
I would like to express my special thanks to Philip Rodrigues who provided insight and expertise that greatly assisted the research. I would also like to  thank Satoshi Nakamura, Lukas Koch and Daniel Cherdack for their comments that greatly improved the results and manuscript. \\

I am extremely grateful to Dave Wark, Kevin Mc Farland, and FNAL for being a constant source of encouragement and support.
\end{acknowledgments}
\appendix
\section{}
The V-A helicity amplitudes of MK model is defined in reference \cite{MK}, however the helicity amplitudes of vector current are given in \Cref{res_HV} for convenience.
\begin{table*}
\centering
\caption{Vector helicity amplitudes of resonant interaction. }
\label{res_HV}
\renewcommand{\arraystretch}{1.3}
\begin{ruledtabular}
 \begin{tabular}{c|c|c|c}
  $\lambda_2$& $\lambda_1$ &$\tilde{F}_{\lambda_2 \lambda_1}^{e_{L}}(\theta, \phi)$& $\tilde{F}_{\lambda_2 \lambda_1}^{e_{R}}(\theta, \phi)$\\[4pt]
 \hline
 $\begin{aligned}
&\scalebox{0.001}{~}\\&\frac{1}{2}\\[7pt] -&\frac{1}{2} \\[7pt] &\frac{1}{2}\\[7pt] -&\frac{1}{2}
\end{aligned}$
&
 $\begin{aligned}
&\scalebox{0.001}{~}\\&\frac{1}{2}\\[7pt] &\frac{1}{2} \\[7pt] -&\frac{1}{2}\\[7pt] -&\frac{1}{2}
\end{aligned}$
&
 $\begin{aligned}
&\scalebox{0.001}{~}\\
\pm&\sum_j\frac{2j+1}{\sqrt{2}} \mathcal{D}^j(R) ~f^V_{-3}(R(I,j=l\pm\frac{1}{2}))~  d^j_{\frac{3}{2} \frac{1}{2}}(\theta) e^{-2i\phi}\\
&\sum_j\frac{2j+1}{\sqrt{2}} \mathcal{D}^j(R)~ f^V_{-3}(R(I,j=l\pm\frac{1}{2}))~  d^j_{\frac{3}{2} -\frac{1}{2}}(\theta)e^{-i\phi}\\
\mp&\sum_j\frac{2j+1}{\sqrt{2}} \mathcal{D}^j(R)~ f^V_{-1}(R(I,j=l\pm\frac{1}{2}))~  d^j_{\frac{1}{2} \frac{1}{2}}(\theta)e^{-i\phi}\\
-&\sum_j\frac{2j+1}{\sqrt{2}} \mathcal{D}^j(R)~ f^V_{-1}(R(I,j=l\pm\frac{1}{2}))~  d^j_{\frac{1}{2} -\frac{1}{2}}(\theta)\\
\end{aligned}$
&
 $\begin{aligned}
 &\scalebox{0.01}{~}\\
-&\sum_j\frac{2j+1}{\sqrt{2}} \mathcal{D}^j(R)~ f^V_{-1}(R(I,j=l\pm\frac{1}{2}))~  d^j_{\frac{1}{2} -\frac{1}{2}}(\theta)\\
\pm&\sum_j\frac{2j+1}{\sqrt{2}} \mathcal{D}^j(R)~ f^V_{-1}(R(I,j=l\pm\frac{1}{2}))~  d^j_{\frac{1}{2} \frac{1}{2}}(\theta)e^{i\phi}\\
-&\sum_j\frac{2j+1}{\sqrt{2}} \mathcal{D}^j(R)~ f^V_{-3}(R(I,j=l\pm\frac{1}{2}))~  d^j_{\frac{3}{2} -\frac{1}{2}}(\theta)e^{i\phi}\\
\pm&\sum_j\frac{2j+1}{\sqrt{2}} \mathcal{D}^j(R) ~f^V_{-3}(R(I,j=l\pm\frac{1}{2}))~  d^j_{\frac{3}{2} \frac{1}{2}}(\theta) e^{2i\phi}
\end{aligned}$
\\\hline
&  &$\tilde{F}_{\lambda_2 \lambda_1}^{e_{-}}(\theta, \phi)$& $\tilde{F}_{\lambda_2 \lambda_1}^{e_{+}}(\theta, \phi)$\\[4pt]
 \hline
 $\begin{aligned}
&\scalebox{0.001}{~}\\&\frac{1}{2}\\[7pt] -&\frac{1}{2} \\[7pt] &\frac{1}{2}\\[7pt] -&\frac{1}{2}
\end{aligned}$
&
 $\begin{aligned}
&\scalebox{0.001}{~}\\&\frac{1}{2}\\[7pt] &\frac{1}{2} \\[7pt] -&\frac{1}{2}\\[7pt] -&\frac{1}{2}
\end{aligned}$
&
 $\begin{aligned}
&\scalebox{0.05}{~}\\
&\frac{|\mathbf k|}{\sqrt{-k^2}}\sum_j\frac{2j+1}{\sqrt{2}} \mathcal{D}^j(R)~ f^{V(-)}_{0+}(R(I,j=l\pm\frac{1}{2}))~  d^j_{\frac{1}{2} \frac{1}{2}}(\theta)e^{-i\phi}\\
\pm&\frac{|\mathbf k|}{\sqrt{-k^2}}\sum_j\frac{2j+1}{\sqrt{2}} \mathcal{D}^j(R)~ f^{V(-)}_{0+}(R(I,j=l\pm\frac{1}{2}))~  d^j_{\frac{1}{2} -\frac{1}{2}}(\theta)\\
\pm&\frac{|\mathbf k|}{\sqrt{-k^2}}\sum_j\frac{2j+1}{\sqrt{2}} \mathcal{D}^j(R)~ f^{V(-)}_{0+}(R(I,j=l\pm\frac{1}{2}))~  d^j_{\frac{1}{2} -\frac{1}{2}}(\theta)\\
-&\frac{|\mathbf k|}{\sqrt{-k^2}}\sum_j\frac{2j+1}{\sqrt{2}} \mathcal{D}^j(R)~ f^{V(-)}_{0+}(R(I,j=l\pm\frac{1}{2}))~  d^j_{\frac{1}{2} \frac{1}{2}}(\theta)e^{i\phi}\end{aligned}$
&
 $\begin{aligned}
 &\scalebox{0.05}{~}\\
&\frac{|\mathbf k|}{\sqrt{-k^2}}\sum_j\frac{2j+1}{\sqrt{2}} \mathcal{D}^j(R)~ f^{V(+)}_{0+}(R(I,j=l\pm\frac{1}{2}))~  d^j_{\frac{1}{2} \frac{1}{2}}(\theta)e^{-i\phi}\\
\pm&\frac{|\mathbf k|}{\sqrt{-k^2}}\sum_j\frac{2j+1}{\sqrt{2}} \mathcal{D}^j(R)~ f^{V(+)}_{0+}(R(I,j=l\pm\frac{1}{2}))~  d^j_{\frac{1}{2} -\frac{1}{2}}(\theta)\\
\pm&\frac{|\mathbf k|}{\sqrt{-k^2}}\sum_j\frac{2j+1}{\sqrt{2}} \mathcal{D}^j(R)~ f^{V(+)}_{0+}(R(I,j=l\pm\frac{1}{2}))~  d^j_{\frac{1}{2} -\frac{1}{2}}(\theta)\\
-&\frac{|\mathbf k|}{\sqrt{-k^2}}\sum_j\frac{2j+1}{\sqrt{2}} \mathcal{D}^j(R)~ f^{V(+)}_{0+}(R(I,j=l\pm\frac{1}{2}))~  d^j_{\frac{1}{2} \frac{1}{2}}(\theta)e^{i\phi}\end{aligned}$
\\
\end{tabular}
\end{ruledtabular}
\end{table*}
\nocite{*}

\bibliography{MK_vector}

\end{document}